\documentclass[draftcls,twocolumn]{IEEEtran}
\usepackage{graphicx}
\usepackage{epsfig}

\ifCLASSINFOpdf
\else
\fi
\begin{document}
%
\title{Effective-mass model of surface scattering in locally oxidized
Si nanowires}
%
%
%

\author{P.~Drouvelis and G.~Fagas 
\thanks{P.~Drouvelis is with the
Institute for Interdisciplinary Scientific Computing, University of Heidelbreg,
Im Neuenheimer Feld 368, 69120, Germany.}
\thanks{G.~Fagas is with the
Tyndall National Institute, Lee Maltings, Cork, Ireland.}
}

%
%

\markboth{ULIS 2009 - CONFERENCE PROCEEDINGS}%
{Drouvelis and Fagas:Scattering model for locally oxidized
Si nanowires}
%



\maketitle

\begin{abstract}
We present a simple model to describe the lowest-subbands surface scattering
in locally oxidized silicon nanowires grown in the [110] direction.
To this end, we employ an atomistically scaled effective mass model
projected from a three-dimensional effective mass equation and apply
a quantum transport formalism to calculate the conductance
for typical potential profiles. Comparison of our results with hole-transport
calculations using atomistic models in conjuction with Density Functional Theory (DFT)
points to an intra-subband scattering mechanism from a potential well.
\end{abstract}

\begin{IEEEkeywords}
Intraband scattering, silicon nanowires, oxidation
\end{IEEEkeywords}

\date{today}

%
\IEEEpeerreviewmaketitle

\section{Introduction}
%
%
%
%

Recent bottom-up~\cite{Yang06} and top-down~\cite{Lansbergen08,Colinge07}
demonstrations of Si channels with diameters of only a few nanometers
shape the roadmap towards the ultimate limits of scaling. Silicon nanowires (SiNWs), in particular,
can be easily integrated with existing semiconductor technologies. As experimental results remain
difficult to obtain and analyze at this lengthscale, modelling has a crucial role to play in
assessing this technology.

Owing to computational and methodological advances, great progress has been achieved
in atomic-scale transport calculations, first in molecular systems~\cite{Cuniberti,Datta}
and later in semiconductor nanowires~\cite{Klimeck,Neophytou2,Luisier,Giorgos}.
Nevertheless, semiconductor physics is most easily captured within the effective mass theory and
many device simulators are built based on this approximation~\cite{Asenov,Goodnick}. This implies that simple models
describing the basic mechanisms are required~\cite{Neophytou1}.
Recently, using an effective mass model we were able to determine the important parameters in
scattering from neutral P-dopants in SiNWs and reproduce results from a first-principles approach within DFT~\cite{Panos}.

Here, in a similar fashion, we take a first step in explaining scattering from Si atoms at the wire
surface promoted to different oxidation states (typically Si$^{+1}$ and Si$^{+2}$).
Due to oxidation, this type of oxygen-related defects are unavoidable at the surface of a pure
semiconductor wire core even in surface treated SiNWs and can be considered as a source of roughness on
the atomic scale~\cite{Luisier,Pantelides}. Using an effectively one-dimensional model of scattering from a well we fit 
the conductance of ultrathin ($<$ 3nm in diameter) SiNWs as obtained from a combined DFT-Green's
function formalism.
An atomic-scale analysis based on calculations from first-principles will be presented elsewhere~\cite{Giorgos}.
This approach allows us to identify the scattering mechanism as intra-subband scattering
from a shallow potential well.

In Section \ref{Sec2}, we summarize the equations of the effective mass model
solved using a quantum transport scattering approach.
In Section \ref{Sec3}, we present a discussion of our results for various typical
scenarios of local surface oxidation. We conclude with Section \ref{Sec4}.

\begin{figure}[h]
\centering
\includegraphics[angle=0,width=4cm]{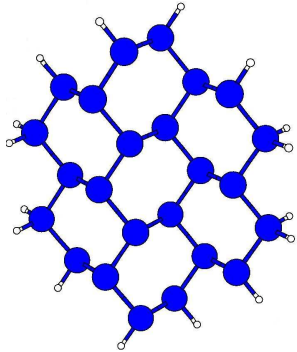}%
\label{Fig1}
\caption{The cross-section of [110]-oriented hydrogenated SiNWs with diameter W = 1.15nm as measured from the distance between furthest Si atoms in the slice.}
\end{figure}

\section{Effective mass equations}
\label{Sec2}

The general Hamiltonian that governs the three-dimensional electron motion reads

\begin{equation}
H =
\frac{\hbar^2}{2} \mathbf{k}^T \stackrel{\textstyle\leftrightarrow}{m}^{-1}_D \mathbf{k}
+ W(x,y,z)
\label{eq1_c1}.
\end{equation}

Here, $\stackrel{\textstyle\leftrightarrow}{m}^{-1}_D$ is the
effective mass tensor in the coordinate system of the device (i.e., x-axis is oriented along the channel)
, $\mathbf{k} = (k_x,k_y,k_z) = -i (\frac{\partial}{\partial x},\frac{\partial}{\partial y},
\frac{\partial}{\partial z})$ is the momentum operator for a fully spatially
quantized electron motion and $W(x,y,z)$ is the single-particle operator for the electrostatic
potential that the electron feels when injected in the channel. In what follows we focus on hole
transport and take that the effective mass tensor is diagonal within our SiNW coordinate system.
This leads to an effective mass Hamiltonian \ref{eq1_c1} for the nanostructure
reading~\cite{Lundstrom}

\begin{equation}
H =
- \frac{\hbar^2}{2}
\sum_{r=x,y,z}\frac{\partial}{\partial r}\frac{1}{m^*_r(x,y,z)} \frac{\partial}{\partial r}
+ W(x,y,z).
\label{eq2_c1}
\end{equation}

\noindent
We also assume that the effective mass varies only along the channel, i.e., the x-direction,
allowing us to use an ansatz of the form

\begin{equation}
\Psi(x,y,z) = \sum\limits_{nm} \phi_{nm}(x) \xi_{nm}(x,y,z).
\label{eq3_c1}
\end{equation}

\noindent
Applying this ansatz and using the orthonormality property for the
function $\xi_{nm}(x,y,z)$ together with a 'smooth disorder' assumption
along transport, i.e. very slow variation, then it is eligible to assume
that the function $\xi_{nm}(x,y,z)$ is independent of $x$. This approximation has
also been used in Ref.~\cite{Wang}. Under these assumptions we end up
with an one-dimensional effective mass equation governing transport within the nanowire

\begin{equation}
\left \{- \frac{\hbar^2}{2} \frac{\partial}{\partial x} \frac{1}{m^*_x(x)} \frac{\partial}{\partial x} +
E_{nm}(x) \right \} \phi_{nm}(x)  = 
E \phi_{nm}(x),
\label{eq4_c1}
\end{equation}

\noindent
where $E_{nm}(x)$ is determined by the solution of

\begin{eqnarray}
\left \{- \frac{\hbar^2}{2}
\sum_{r=x,y} \frac{1}{m^*_r(x)} \frac{\partial^2}{\partial^2 r}
+ W(y,z;x)
\right \}
\xi_{nm}(x,y,z) = \\ \nonumber
E_{nm}(x) \xi_{nm}(x,y,z)
\label{eq5_c1}
\end{eqnarray}

The quantum mechanical scattering problem corresponding to Equation \ref{eq4_c1} is then
solved to yield the transmission coefficient, $T_i$, for each channel and, hence,
the Landauer conductance $G = (2e^2/h) T = (2e^2/h) \sum T_i$.

\section{Results}
\label{Sec3}
For definitiveness, we consider SiNWs grown along the [110]-direction and
fix the diameter to $W = 1.15$nm. Similar results are obtained for the lowest
subbands in wires with larger W. Figure 1 shows the cross-section of the wire
channel. The atomistic model for the promotion of Si atoms to the states Si$^{+1}$ and Si$^{+2}$
consists of starting from the neutral oxidation state Si$^{0}$ at the surface of the hydrogenated wires
and introducing oxygens as shown in Figure 2. The resulting Si-O-H and Si-O-Si bonds
are the simplest cases of oxygen bridge- and back- bonds.

\begin{figure}[h]
\centering
\includegraphics[angle=0,width=4cm]{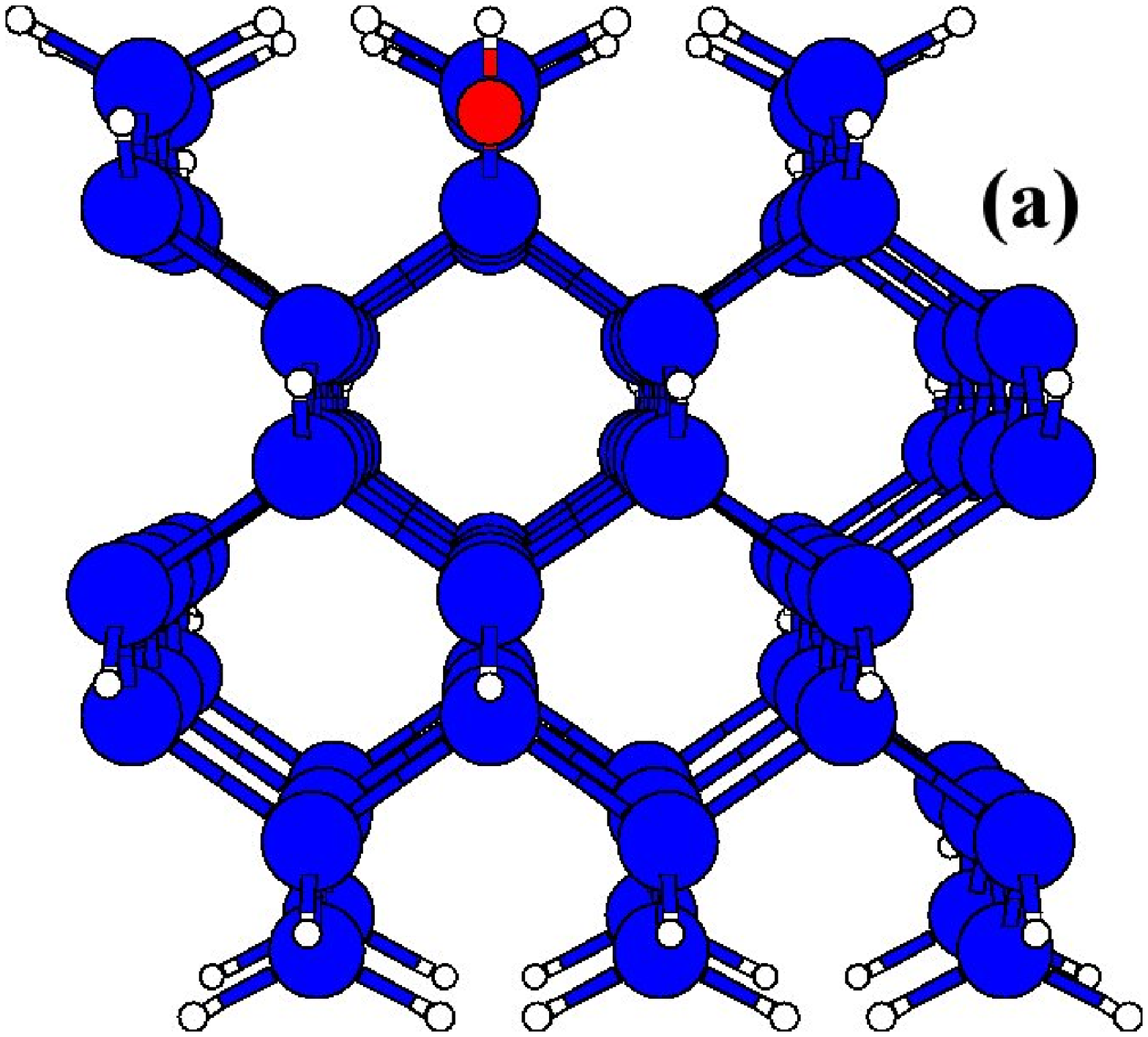}%
\vspace{1cm}
\centering
\includegraphics[angle=0,width=4cm]{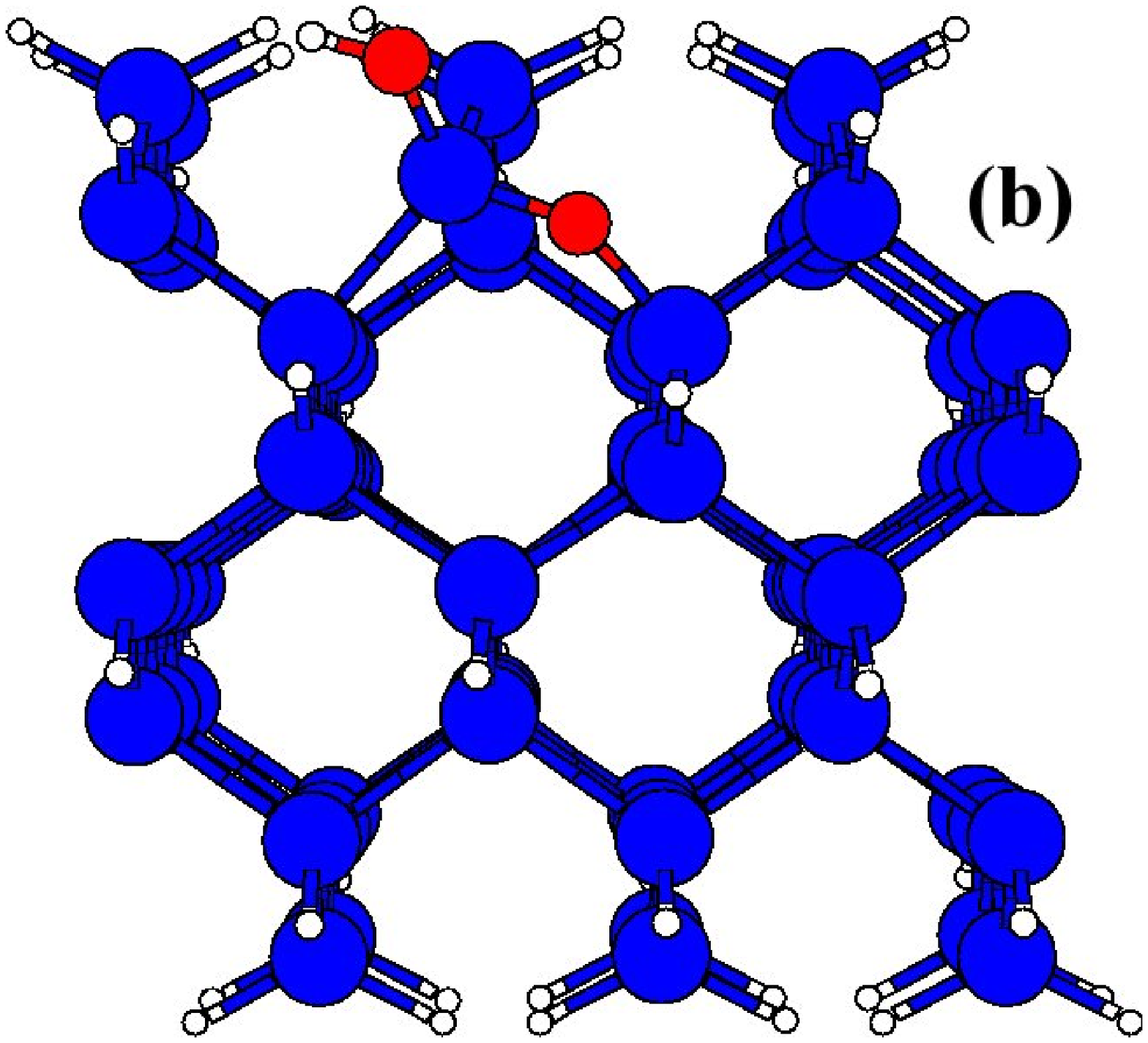}%
\label{Fig2}
\caption{Examples of an (a) Si-O-H bond, (b) Si-O-Si back-bond. The oxidized surface Si atom
is at the Si$^{+1}$ and Si$^{+2}$ oxidation state, respectively.}
\end{figure}

Before performing a transport calculation using Eq. \ref{eq4_c1} it is instructive to
compare the electronic structure of the pristine hydrogenated wire with the bands
of a large enough supercell that includes an oxygen-derived defect. This is done
in Figure 3, where we plot the bands derived from a density functional tight-binding (DFTB)
approach. The latter is an efficient approximation to the Kohn-Sham DFT scheme.
We first note that the effective mass as determined by the curvature of the bands
does not change by the introduction of the oxygen. In general, impurities cause
variations of the effective mass $m^*_x(x)$ along the channel. But here this effect
can be disregarded.

\vspace{0.5cm}
\begin{figure}[h]
\centering
\includegraphics[angle=0,width=7cm]{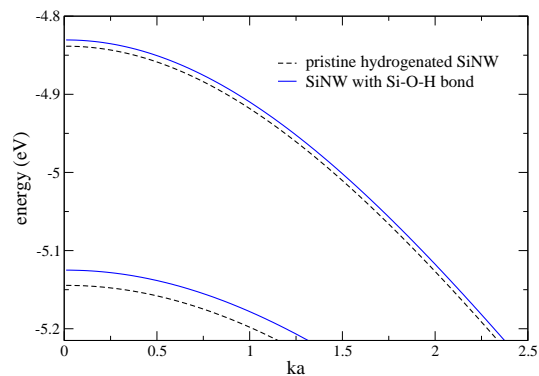}%
\label{Fig3}
\caption{Band structure of [110]-oriented Si nanowire with cross-section as in Figure 1. The supercell lattice
parameter a is 1.92nm.}
\end{figure}

A second observation is that alignment of the bands in a pristine-oxidised-pristine wire arrangement
would yield a potential well of the order of a few meV for hole propagation in both subbands.
For the effective mass model, we take this into account as a phenomenological parameter
in Eq. \ref{eq4_c1} through the x-dependence of W. The conductance for
a potential well extending over a region of length $L \sim 1 nm$ is plotted in Figure 4.
Also in Figure 4, we show for comparison the result of a combined DFTB-Green's
function approach as well as the ideal hole conductance of a pristine hydrogenated wire.
It is evident that for both cases of an Si-O-H bond and an Si-O-Si back-bond
the simple model of considering the region around the oxidized Si site as
a shallow potential well yields a remarkable quantitative agreement.
This picture is consistent with ballistic transport and the positive oxidation state of Si.
We note that within the effective mass model only intrasubband scattering is considered,
thereby, implying that this is the dominant scattering mechanism. As an aside
we confirmed this by independent analysis of the quantum mechanical S-matrix in the atomic-scale
model.

\begin{figure}[h]
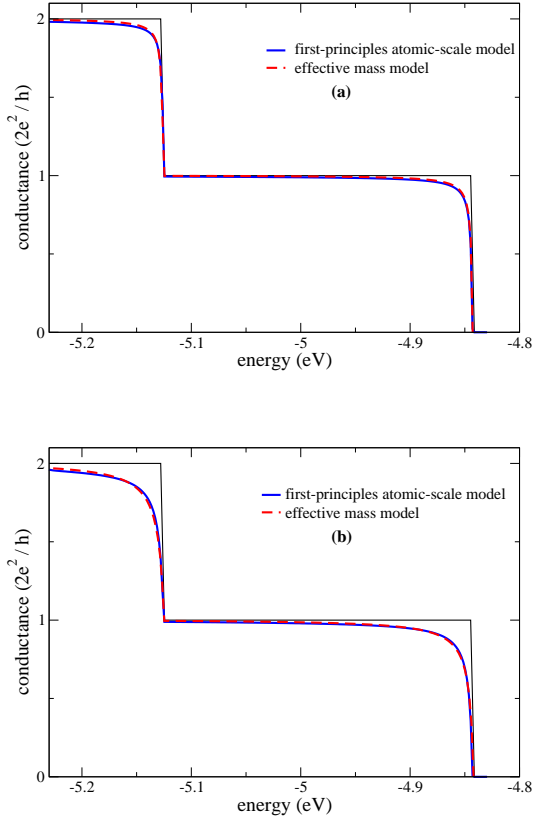

\centering
\includegraphics[angle=0,width=7cm]{Fig4a}%
\vspace{1cm}
\centering
\includegraphics[angle=0,width=7cm]{Fig4b}%
\label{Fig4}
\caption{Landauer conductance of locally oxidized SiNWs corresponding to the
oxygen-related defects of Figure 2; models as indicated.}
\end{figure}

\section{Conclusion}
\label{Sec4}
In conclusion, effective mass models are useful in analyzing measurements, model building
and for practical calculations of charge carrier transport in semiconductor wires.
To this end, we derived a simple picture within this framework to explain results of
hole-transport calculations in locally oxidized silicon nanowires grown in the [110] direction
using atomic-scale models in conjuction with Density Functional Theory (DFT).
This appoach points to an intra-subband scattering mechanism from a potential well
and may further apply in other impurity systems.


%



\section*{Acknowledgment}
P.~Drouvelis acknowledges financial support from the
Irish Research Council for Science Engineering and Technology
while at Tyndall National Institute where part of this work
was performed. The research of G.~Fagas is supported by
Science Foundation Ireland.

\ifCLASSOPTIONcaptionsoff
  \newpage
\fi

\end{document}